\documentclass{article}
\usepackage{smc}
\usepackage{times}
\usepackage{ifpdf}
\usepackage[english]{babel}
\usepackage{cite}
\usepackage{tabularx}
\usepackage{tablefootnote}
\usepackage{booktabs}
\usepackage{xcolor}
\def\papertitle{PAPER TEMPLATE FOR SMC 2022}
\def\firstauthor{Xiaoxue Gao}
\def\secondauthor{Chitralekha Gupta}
\def\thirdauthor{Haizhou Li}
\newif\ifpdf
\ifx\pdfoutput\relax
\else
   \ifcase\pdfoutput
      \pdffalse
   \else
      \pdftrue
\fi

\ifpdf 
  \usepackage[pdftex,
    pdftitle={\papertitle},
    pdfauthor={\firstauthor, \secondauthor, \thirdauthor},
    bookmarksnumbered, 
    pdfstartview=XYZ 
   ]{hyperref}

  \usepackage[pdftex]{graphicx}
  \graphicspath{{./figures/}}
  \DeclareGraphicsExtensions{.pdf,.jpeg,.png}

  \usepackage[figure,table]{hypcap}

\else 
  \usepackage[dvips,
    bookmarksnumbered, 
    pdfstartview=XYZ 
  ]{hyperref}  

  \usepackage[dvips]{epsfig,graphicx}
  \graphicspath{{./figures/}}
  \DeclareGraphicsExtensions{.eps}

  \usepackage[figure,table]{hypcap}
\fi
\hypersetup{
    colorlinks,%
    citecolor=black,%
    filecolor=black,%
    linkcolor=black,%
    urlcolor=black
}

\title{Music-robust Automatic Lyrics Transcription of Polyphonic Music}

\threeauthors
  {Xiaoxue Gao} {{National University of Singapore,}\\ {Singapore} \\
     {\tt \href{xiaoxue.gao@u.nus.edu}{xiaoxue.gao@u.nus.edu}}}
  {Chitralekha Gupta} {{National University of Singapore,}\\
     {\tt \href{chitralekha@nus.edu.sg}{chitralekha@nus.edu.sg}}}
  {Haizhou Li} {{The Chinese University of Hong Kong, Shenzhen, China}\\
     {\tt \href{haizhou.li@cuhk.edu.cn}{haizhou.li@cuhk.edu.cn}}}
     
\threeauthors
  {Xiaoxue Gao} {{National University of Singapore,}\\
  {Singapore} \\
     {\tt \href{xiaoxue.gao@u.nus.edu}{xiaoxue.gao@u.nus.edu}}}
  {Chitralekha Gupta} {{National University of Singapore,}\\
  {Singapore} \\
     {\tt \href{chitralekha@nus.edu.sg}{chitralekha@nus.edu.sg}}}
  {Haizhou Li} {{The Chinese University of Hong Kong,}\\
  {Shenzhen, China} \\
     {\tt \href{haizhou.li@cuhk.edu.cn}{haizhou.li@cuhk.edu.cn}}}
\begin{document}
\capstartfalse
\maketitle
\capstarttrue

\begin{abstract}
Lyrics transcription of polyphonic music is challenging because singing vocals are corrupted by the background music. To improve the robustness of lyrics transcription to the background music, we propose a strategy of combining the features that emphasize the singing vocals, i.e.~\textit{music-removed} features that represent singing vocal extracted features, and the features that capture the singing vocals as well as the background music, i.e.~\textit{music-present} features. We show that these two sets of features complement each other, and their combination performs better than when they are used alone, thus improving the robustness of the acoustic model to the background music. Furthermore, language model interpolation between a general-purpose language model and an in-domain lyrics-specific language model provides further improvement in transcription results. Our experiments show that our proposed strategy outperforms the existing lyrics transcription systems for polyphonic music. Moreover, we find that our proposed music-robust features specially improve the lyrics transcription performance in metal genre of songs, where the background music is loud and dominant.
\end{abstract}

\section{Introduction}
\label{sec:intro}

While significant progress has been achieved in automatic speech recognition (ASR)~\cite{sun2019adversarial,povey2016,hinton2012deep,sainath2013deep,xiong2018microsoft} and deep learning~\cite{gao2019speaker,gao2020personalized}, lyrics transcription of polyphonic music remains unsolved. In recent years, there has been an increasing interest in lyrics recognition of polyphonic music, which has potential in many applications such as the automatic generation of karaoke lyrical content, music video subtitling, query-by-singing \cite{mesaros2013singing} and singing processing~\cite{gao2018nus,sharma2021nhss,vijayan2018analysis}. The goal of lyrics transcription of polyphonic music is to recognize the lyrics from a song that contains singing vocals mixed with background music.

The main challenge for a lyrics transcription system is that the background music interferes with the singing vocals, thereby degrading the lyrics intelligibility. Past studies have tackled the background music interference with broadly two approaches: 1) by incorporating singing vocal extraction (a \textit{music removal} approach) \cite{gupta2019,mesaros2010automatic,dzhambazov2015modeling} as a pre-processing module; and 2) by using the background music knowledge to enhance the model (a \textit{music-present} approach)\cite{stoller2019,gupta2019automatic}.

In the \textit{music-removal} approach, various singing vocal separation techniques have been studied to suppress the background music and extract the singing vocals from the polyphonic music for acoustic modeling \cite{gupta2019,mesaros2010automatic,dzhambazov2015modeling,fujihara2011lyricsynchronizer}. 
However, due to imperfection in music removal as well as the distortions associated with the inversion of a magnitude spectral representation, the extracted time-domain singing vocal signals often contain artifacts. Acoustic model trained on such extracted vocals are far from perfect~\cite{gupta2019,gupta2019automatic}. Another \textit{music-removal} approach is to train acoustic model only on clean singing vocals, and use vocal extraction on test polyphonic music signals at the time of inference~\cite{guptalyrics,dabikesheffield,demirel2021low}. However, singing-only acoustic models need to be trained using a large amount of solo singing data~\cite{demirel2020automatic}, and the two-step procedure suffers from mismatch between the acoustic features between training and testing, thereby causing degradation of the performance of acoustic modeling in lyrics recognition. 

Rather than directly applying a solo-singing acoustic model to polyphonic data, polyphonic audio adaptation~\cite{gupta2019acoustic} techniques are used to adapt a model trained on a large amount of solo singing data with a small amount of in-domain polyphonic audio files, and this in-domain adaptation is found to be outperforming the solo-singing acoustic models adapted with extracted singing vocals. This suggests the polyphonic data, i.e.~singing vocals+background music, helps in learning the spectro-temporal variations of the background music more than the extracted vocals.

In the \textit{music-present} approach, instead of removing the background music, the system is directly trained with the polyphonic music input for lyrics recognition~\cite{stoller2019,gupta2019automatic} . For example, Stoller et al.~\cite{stoller2019} used an end-to-end wave-U-net model to predict character probabilities directly from the polyphonic audio, while Gupta et al.~\cite{gupta2019automatic} used a Kaldi-based acoustic model training with polyphonic music conditioned with music genre related information. These studies show that the task of lyrics acoustic modeling can benefit from the knowledge of background music.

\begin{figure*}[t]
\centering
\flushleft 
\includegraphics[width=1.99\columnwidth]{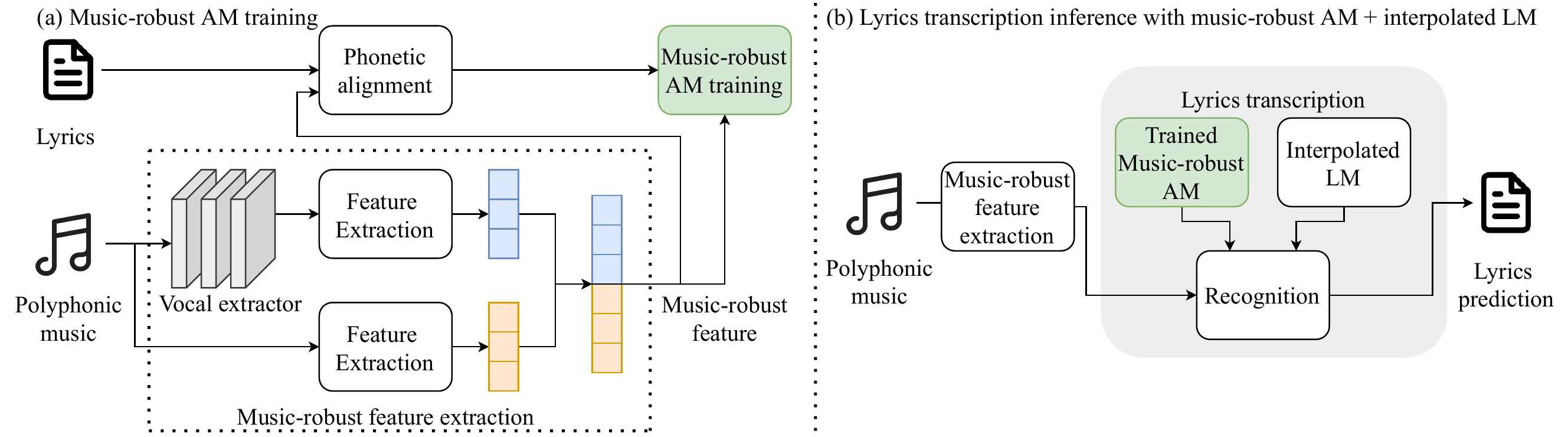}
\caption{Model architecture of the proposed music-robust lyrics transcription framework using the music-robust AM and the interpolated LM.} 
\label{fig}
\end{figure*}
For speech recognition~\cite{feng2014speech} in noisy environment, the noise-robust speech recognition techniques include both speech enhancement techniques (that suppress the background noise)~\cite{tu2018hybrid,tu2019dnn,hermus2006review} and training with environment-matched noisy data~\cite{sharma2000feature,seltzer2013investigation,zhang2018deep}. Inspired by the success of noise-robust speech recognition, we propose a combination of \textit{music-removed} features and \textit{music-present} features to build a \textit{music-robust} automatic lyrics transcription (ALT) system. We believe that this combination of features will provide complementary information that will be helpful for lyrics transcription, i.e.~music-removed features will focus on the lyrical content, while the music-present features will compensate for the distortions caused by the imperfect vocal extraction, and that is the focus of this paper.  

Furthermore, we investigate the interpolation between a high resource speech corpus language model with an in-domain song-lyrics language model for capturing the domain-specific semantics, which yields a better performing system for the task of lyrics transcription. 

The main contributions of this paper are: 
\begin{itemize}
\item  We investigate different singing vocal-related features for acoustic modeling in automatic lyrics transcription of polyphonic music;
\item  We propose a novel music-robust feature which is designed to enhance vocal-specific lyrical information and compensate for the corrupted temporal-spectro vocal components caused by singing vocal extractors; 
\item  By combining the music-present features with the music-removed features, we show that lyrics intelligibility is enhanced in the presence of background music.
\end{itemize}

The rest of this paper is organized as follows. In Section~\ref{Singing Representation based Lyrics Transcription}, we will present the proposed music-robust approach. In Section~\ref{Experiments}, the experimental setting and baselines will be presented. In Section~\ref{Results}, we will discuss the experiments results. We will summarize the contributions of this paper in Section~\ref{Conclusion}.

\section{Music-robust automatic Lyrics Transcription}

\label{Singing Representation based Lyrics Transcription}
We propose a music-robust automatic lyrics transcription framework for polyphonic music, which includes a music-robust acoustic model (AM) and an interpolated language model (LM).

One way of performing lyrics transcription in polyphonic music is to extract singing vocals via source separation approaches~\cite{hennequin2019spleeter,stoter2019open,defossez2019music,stoller2018wave} for acoustic modeling. Features extracted from music-removed singing vocals focus on the vocals that contain lyrical information. However, singing vocals extraction systems are not perfect~\cite{hennequin2019spleeter,stoter2019open,defossez2019music,stoller2018wave}, and the extracted vocals suffer from the distortions caused by incomplete or distorted vocal feature prediction, as well as imperfect time-domain signal reconstruction from the estimated magnitude spectrogram. As a result, the extracted music-removed acoustic features  may lose important lyrics-related information that will affect lyrics transcription performance. The polyphonic signal, on the other hand, has unaltered vocals information. However, polyphonic music often contains loud background music that overpowers the singing vocals~\cite{condit2015catching}. With polyphonic features alone, the lyrics transcription performance will get affected due to the background music. 

We hypothesize that a combination of polyphonic features (or music-present features) and extracted vocal features (or music-removed features) will highlight the vocal-related information while compensating for the distorted parts, thus providing a robust framework for lyrics transcription of polyphonic music.
The proposed framework is illustrated in Fig.~\ref{fig}.

\subsection{Music-robust Feature}
\label{Singing Vocal Representation Learning}
We combine two kinds of features to form the music-robust feature as shown in Fig.~\ref{fig} (a). The first feature is the \textit{music-present} MFCCs which are obtained directly from the mixed polyphonic input audio (shown in orange in Fig.~\ref{fig} (a)). The second feature is the \textit{music-removed} MFCCs (shown in blue in Fig.~\ref{fig} (a)) which are obtained from the singing-only vocals extracted from the polyphonic mixture input audio through a singing vocal extractor. The music-robust feature is formed by concatenating the music-removed MFCC feature with the music-present MFCC feature. 

\subsection{Music-robust Acoustic Model} 
We establish the inter-relations of the music-present feature and music-removed feature through the deep neural networks in a data-driven manner. These two sets of features are stacked together as the inputs to a widely used standard ASR system that consists of an acoustic model (AM) which is trained by a factorized time-delay neural network (TDNN-F)~\cite{povey2018}.

\subsection{Interpolated Language Model}
Prior work have explored the language models directly trained on speech corpus text~\cite{gupta2019automatic} or lyrics text~\cite{gupta2019automatic,demirel2021low,mesaros2010automatic,stoller2019} for lyrics transcription of polyphonic. These approaches only benefit from one domain, and the cross-domain leveraging impact of linguistic contents can be explored further.
In order to make better use of high resource speech text and adapt the linguistic peculiarities of lyrics of songs such as connecting words, grammar and rhythmic patterns \cite{fang2017discourse} to speech text, we propose to construct an interpolated language model (LM) that bridges between a small in-domain lyrics LM trained on sung lyrics text corpus and a general LM trained on a large vocabulary speech corpus text for lyrics transcription. 

We present the inference stage of music-robust automatic lyrics transcription using the proposed music-robust AM and interpolated LM in Fig.~\ref{fig} (b).
Given a piece of polyphonic music (background music + singing vocal), the music-robust features are extracted (Section~\ref{Singing Vocal Representation Learning}), and with the help of the trained music-robust AM and interpolated LM, lyrics predictions are obtained by feeding the music-robust features to music-robust AM.
\section{Experiments}
\label{Experiments}
\subsection{Datasets}
As shown in Table \ref{tab:datasets}, the training data for acoustic modeling consist of DALI \cite{meseguer2018dali} and a proprietary dataset from NUS of 517 popular English songs. DALI has 3,913 English polyphonic audio tracks\footnote{There are a total of 5,358 audio tracks in DALI, where only 3,913 English audio links were accessible from Singapore.} and comprises of 180,034 line-level audio and lyrics transcription with a total duration of 208.6 hours. The songs in the NUS proprietary dataset was split into lines automatically using the system in \cite{gupta2019automatic}. This dataset consists of 26,462 line-level audio and lyrics transcription with a total duration of 27.0 hours. We also used 100 songs from DALI dataset \cite{gupta2019automatic} which are not present in its training dataset, as a development set. We fine-tune our language model on this dev set.

We evaluate the performance of our proposed lyrics transcription framework on three widely used polyphonic test datasets -- Hansen\footnote{Total 9 songs including the song ``clock'' with corrected version for the errors in the ground-truth lyrics transcription.}\cite{hansen2012recognition}, Mauch\cite{mauch2010lyrics}, and Jamendo\cite{stoller2019}. The test datasets were English songs obtained from the respective authors for our research.

\begin{table}[t]
\caption{Dataset description.}
\label{tab:datasets}
\begin{tabular}{c|rrr}
\toprule
\textbf{Name} & \textbf{\# songs} & \textbf{\# lines} & \textbf{Total duration}\\ \midrule
DALI \cite{meseguer2018dali} & \begin{tabular}[c]{@{}c@{}}3,913 \end{tabular}& \begin{tabular}[c]{@{}c@{}}180,034 \end{tabular} &\begin{tabular}[c]{@{}c@{}}208.6 hours\end{tabular}\\ 
NUS & 517 & \begin{tabular}[c]{@{}c@{}}26,462 \end{tabular} & 27.0 hours\\

DALI-dev \cite{meseguer2018dali} & 100 & \begin{tabular}[c]{@{}c@{}}5,356 \end{tabular}& 3.9 hours\\
\bottomrule
\end{tabular}%
\end{table}
\subsection{Acoustic Model and Singing Vocal Extraction}
We employ the open-unmix system~\cite{stoter2019open} for singing vocal extraction. Open-unmix is trained by parallel polyphonic mixture and clean singing vocal audio using bidirectional LSTM, and it learns to predict the magnitude spectrogram of singing vocal from the magnitude spectrogram of the corresponding mixture inputs (singing vocal+background music). It was the state-of-the-art open-source music source separation system in the source separation challenge SiSEC 2018 \cite{sisec2018}. We use the pre-trained open-unmix model ``umx'' published in~\cite{stoter2019open}.

The ASR system used in these experiments is trained using the Kaldi ASR toolkit~\cite{povey2011kaldi}.
The state-of-the-art Kaldi acoustic modeling pipeline consists of a context dependent phonetic alignment model, to get time-alignments between the lyrics and the audio, and an acoustic modelling network. The phonetic alignment model is GMM-HMM based trained with 40k Gaussians using MFCC features. The frame rate and length are 10 and 25 ms, respectively. For acoustic modeling, the state-of-the-art factorized time-delay neural network (TDNN-F) architecture~\cite{povey2018} is employed. 

The possible input features are as follows:
\begin{itemize}
\item \textbf{Poly}: AM is trained with a 40-dimensional high-resolution MFCC from polyphonic audio as input features. The phonetic alignment model input features are 39 dimensional MFCC features including the deltas and delta-deltas from polyphonic music. We refer to these poly features as \textit{music-present} features. 

\item \textbf{Vocal}: AM is trained with a 40-dimensional high-resolution MFCCs from extracted vocal training set obtained by open-unmix separation~\cite{stoter2019open} as input features. The phonetic alignment model input features are 39 dimensional MFCC features including the deltas and delta-deltas from extracted vocal. We refer to these features as \textit{music-removed} features. 

\item \textbf{Music-robust}: AM is trained with a 40-dimensional extracted vocal high-resolution MFCCs (music-removed features) stacked with the 40-dimensional polyphonic high-resolution MFCCs (music-present features) as input features. Similarly, the phonetic alignment model input features are 39 dimensional MFCC features from Vocal model stacked with 39 dimensional MFCC features from Poly model.
\end{itemize}

The training data was augmented with speed permutation of the audio files by reducing (x0.9) and increasing (x1.1) the speed of each utterance~\cite{ko2015}. 
A duration-based modified pronunciation lexicon is used that is detailed in~\cite{gupta2018automatic}.
For training the acoustic model, a frame subsampling rate is set to 3 providing an effective frame shift of 30 ms. The minibatch size is 64 where the data chunks of variable sizes of 150, 110, 100 are processed in each minibatch. Time-delay factorized neural network (TDNN-F) is used for AM which is trained for 3 epochs using Stochastic Gradient Descent (SGD) optimizer with initial and final learning rate of 0.00015 and 0.000015, respectively. TDNN-F~\cite{povey2016} consists of sixteen hidden layers with  1,536 dimensions and a bottleneck dimension of 160, where each layer in the network is followed by a ReLU and batch normalization. All the acoustic models were trained according to the standard Kaldi recipe (version 5.4)~\cite{povey2011kaldi}, where the default setting of hyperparameters provided in the standard recipe was used and no hyperparameter tuning was conducted during the acoustic model training.

\begin{table}[t]
\centering
\caption{Comparison of lyrics recognition (WER\%) performance with different language models on Poly model.}

\footnotesize
\begin{tabular}{c|ccc}
\toprule
        & General LM & Lyrics LM & Inter LM       \\ \midrule
Hansen & 62.69      & 51.78     & \textbf{50.95} \\ 
Jamendo & 61.56      & 57.57     & \textbf{56.81} \\ 
Mauch   & 53.84      & 44.18     & \textbf{43.65} \\ \bottomrule
\end{tabular}
\label{LM results}
\end{table}

\subsection{Language Model}

The interpolated language model was built with the following steps: 1) lyrics LM was trained on the training corpus that contains the lyrics of songs; 2) general LM is trained on a large corpus of spoken English text; and 3) interpolated LM was built by interpolating between lyrics LM and general LM. 
\begin{table}[t]

\caption{Comparison of lyrics recognition (WER\%) performance with different singing vocal-related features using the proposed interpolated LM.}

\centering
\footnotesize
\begin{tabular}{c|ccc}
\toprule
        & Poly  & Vocal & Music-robust   \\ \midrule
Hansen  & 50.95 & 52.42 & \textbf{48.33} \\ 
Jamendo & 56.81 & 59.22 & \textbf{55.73} \\ 
Mauch   & 43.65 & 45.37 & \textbf{41.23} \\ \bottomrule
\end{tabular}
\label{model results}
\end{table}

The in-domain lyrics LM is built using the lyrics corpus of the songs in training datasets (Table~\ref{tab:datasets}). The general LM is a 3-gram ARPA LM, pruned with threshold 3e-7 obtained from the open source of LibriSpeech language models \footnote{http://www.openslr.org/11/}. The interpolated LM is built with the interpolation weight that yields the lowest perplexity on the DALI development set (Table \ref{tab:datasets}). The interpolated LM uses the standard 3-grams with interpolated Kneser-Ney smoothing using SRILM toolkit~\cite{stolcke2002srilm}.

\section{Results}
\label{Results}

We present our experimental results on the influence of the different input features for acoustic modeling and the different language models, and compare the proposed approach with the existing approaches as well as explore the effect of segmentation of the test datasets. Music genre analysis and error analysis under different models are also presented to better understand the results. Word confidence score analysis are presented for analysing the results. To assess the quality of lyrics transcription, we compute word error rate (WER), a standard metric of evaluation of ASR, which is the percentage of the total number of insertions, substitutions, and deletions with respect to the total number of words. 

\label{sec:lm}
\subsection{Language Model}

To compare the performances of different language models, we present recognition performance results on a Poly baseline model, that is trained only on the Poly features, with different language models. As shown in Table~\ref{LM results}, lyrics LM is significantly better than the general LM across all three test datasets. This indicates that the in-domain lyrics semantic components is beneficial to the same domain lyrics transcription task. We further found that the interpolated LM outperforms the lyrics LM in terms of word error rate (WER \%) for all the test datasets in lyrics transcription, which indicates that the low-resource lyrics LM can be enhanced with the help of a large resource of general textual information. We present the rest of the experiments using the proposed interpolated LM.

\subsection{Input Features}

As can be seen in Table~\ref{model results}, system Poly outperforms the system Vocal by 2-5\% showing that the music-present features that preserves the spectro-temporal variations of the vocals can help in transcription performance. Moreover, the proposed music-robust model outperforms other baseline systems across all the three test datasets, and shows 5-6 \% improvement over the Poly model. This indicates that the vocal-specific features and the music-present features can complement each other and the combination of them provides more vocal-aware information that relates to the lyrical contents, therefore the combination of these two features in the music-robust acoustic model is effective in improving the accuracy of lyrics transcription in polyphonic audio. 
\begin{table}[t]
\centering
\caption{Comparison of lyrics recognition (WER\%) performance with the segmented testsets.}

\footnotesize
\begin{tabular}{c|ccc}
\toprule
                & Poly  & Vocal & Music-robust   \\ \midrule
Hansen          & 50.95 & 52.42 & 48.33          \\ 
Hansen-segment  & 45.08 & 45.68 & \textbf{42.36} \\ 
Jamendo         & 56.81 & 59.22 & 55.73          \\ 
Jamendo-segment & 54.58 & 56.50 & \textbf{53.21} \\ 
Mauch           & 43.65 & 45.37 & 41.23          \\ 
Mauch-segment   & 40.56 & 43.05 & \textbf{39.68} \\ \bottomrule
\end{tabular}
\label{segment}
\end{table}

\begin{table}[t]
\centering
\caption{Lyrics transcription performance (WER\%) by music genre on the test set for three models, Vocal, Poly and Music-robust.}

\footnotesize
\begin{tabular}{c|ccc}
\toprule
    \textbf{Statistics}  & \textbf{metal}  & \textbf{pop}   & \textbf{hiphop} \\ \midrule
          \# songs in Hansen &  3&5  & 1 \\ 
     \# songs in Jamendo & 7 &9  &4  \\ 
          \# songs in Mauch &8  &12  &0  \\ \midrule
      \# songs in all testset & 17 & 27 & 5 \\ 
\toprule
  \textbf{Models}    & \textbf{metal}  & \textbf{pop}   & \textbf{hiphop} \\ \midrule
Poly  & 60.64 &37.87  & \textbf{56.89}  \\ 
Vocal &60.79  &39.86  &60.51  \\ 
Music-robust & \textbf{56.60}  & \textbf{36.63} &  58.42\\ \bottomrule
\end{tabular}
\label{genre_results}
\end{table}

\begin{table*}[t]
\centering
\caption{Analysis of lyrics recognition performance in terms of errors (insertions, deletions, and substitutions) for the segmented testsets. These are the number of words in error in each category and the percentage of each error with respect to the total number of words.}

\footnotesize
\begin{tabular}{cccc}
\toprule
                & Poly  & Vocal & Music-robust   \\ \midrule
Hansen-segment  & 98 ins, 307 del, 790 sub & 108 ins, 265 del, 838 sub & 105 ins, 291 del, 727 sub \\ 
Jamendo-segment & 205 ins, 962 del, 1867 sub & 181 ins, 1097 del, 1863 sub & 263 ins, 768 del, 1927 sub \\ 
Mauch-segment   & 153 ins, 636 del, 1165 sub & 156 ins, 700 del, 1218 sub & 147 ins, 674 del, 1091 sub \\\midrule
All   & 456 ins, 1905 del, 3822 sub & 445 ins, 2062 del, 3919 sub & 515 ins, 1733 del, 3745 sub \\
All / \# total words & 3.50\% ins, 14.62 \% del, 29.34\% sub &\textbf{3.42\%} ins, 15.83 \% del, 30.08\% sub & 3.95\% ins, \textbf{13.3\%} del, \textbf{28.75\% }sub \\
\bottomrule
\label{stats_error}
\end{tabular}
\end{table*}

\begin{table*}[t]
\centering
\caption{Comparison of lyrics recognition example with three models. \#csid means the number of correct words (C), substitutions (S), insertions (I) and deletions (D).}

\footnotesize
\begin{tabular}{c|lccc}
\toprule
             Model   & ref  GUESS  THAT'S  WHAT  MAKES  ME  THE  ASS  I  SHOULD'VE   KNOWN & error pattern & \#csid & WER (\%)  \\ \midrule
Poly  & hyp    IF   THERE'S  ROOM  MAKES  ME   AS   I   SHOULD     HAVE    KNOWN &S       S       S     C     C   S    S      S        S        C  &3 7 0 0 & 70\\ 
Vocal & hyp   I   GUESS  THAT'S  WHAT  MAKES  ME  BUT   AS  A     SHED    AROUND & I     C       C      C     C     C   S    S   S      S         S & 5 5 1 0 &60\\ 
Music-robust   & hyp    IF   THAT'S  WHAT  MAKES  ME  THAT  ASS  I  SHOULD'VE  KNOWN & S       C      C     C     C    S    C   C      C        C  & 8 2 0 0& 20\\ \bottomrule
\end{tabular}
\label{example}
\end{table*}
\subsection{Genre Analysis}

We analyse the lyrics transcription performance of different music genres in the three test sets, whose genre distribution is summarized in Table~\ref{genre_results} according to the three broad music genre classes -- pop, hiphop and metal, as given in \cite{gupta2019automatic}. 
As shown in Table~\ref{genre_results}, the music-robust system outperforms Poly and Vocal systems for metal and pop songs, and Poly model is superior to Vocal and Music-robust models for hiphop songs. 

This result indicates the complementary nature of the music-present and the music-removed features through the nature of the songs. Metal songs have louder background music than other genres~\cite{condit2015catching}, so the music-removed features have the advantage of clearing up the background ``noise'', while the music-present features compensate for the distortions caused by the vocal extractor. However, hip-hop songs consist of a lot of rap, i.e.~the amount of vocals is high, so the amount of distortions caused by the vocal extractor is also high (Vocal model performs the worst). Although the music-present features compensate a bit for the distortions (music robust is better than vocals only), music-robust model is worse than Poly alone. One should note here that there are only 5 hip-hop songs in the test set, so this observation on hip-hop songs may not be general enough.

\subsection{Error Analysis}
To better analyze the error patterns statistically, we conduct the word error analysis for different models on the segmented testsets in Table~\ref{stats_error} and present a decoded example from different models in Table~\ref{example}. 

Specifically, we compare the substitution, insertion, and deletion errors in the predicted transcriptions of the Poly, Vocal, and Music-Robust systems in Table~\ref{stats_error}. In our proposed music-robust model, the substitution errors decrease for hansen and mauch datasets, and deletion errors decrease for jamendo dataset, compared to the Poly and Vocal baselines. In general, the proposed music-robust system is able to reduce deletion and substitution errors, compared with the Poly and Vocal systems. This implies that the music-robust feature is helpful to amend the distorted vocal parts that come from the vocal extraction process in Vocal system. The overall improvement from Poly system to Music-robust system indicates that the music-robust model provides better vocal-related information that is helpful to lyrics transcription compared with the Poly system. 
We also show an example transcription output for the same audio clip using the three models in Table~\ref{example}. The output transcription of the music-robust model achieves fewer deletion and substitution errors, which further verifies our idea.

\begin{table*}[t]

\centering
\caption{Comparison of lyrics recognition (WER\%) performance with the existing approaches.}

\centering
\footnotesize
\begin{tabular}{c|cccccccc}
\toprule
        & DS~\cite{stoller2019}   & CG~\cite{gupta2019automatic}    & RB1~\cite{dabikesheffield}   & DDA2~\cite{demirel2020recursive}  & DDA3~\cite{demirel2020recursive}  & GGL1~\cite{gaolyrics}  & GGL2~\cite{gaolyrics}   & Music-robust           \\ \midrule
Hansen-ex  & -     & - &   84.98    &  77.12     & 66.70      &     50.88  &    49.85   & \textbf{48.06} \\ 
Jamendo & 77.80 & 59.60 & 86.70 & 72.14 & 73.09 & 60.98 & 62.46 & \textbf{55.73} \\ 
Mauch   & 70.90 & 44.00 & 84.98 & 75.39 & 80.66 & 47.25 & 49.54 & \textbf{41.23} \\ \bottomrule
\end{tabular}
\label{existing}
\end{table*}
\subsection{Word Confidence Analysis}
\begin{figure}[t]
\centering
\includegraphics[width=0.99\columnwidth]{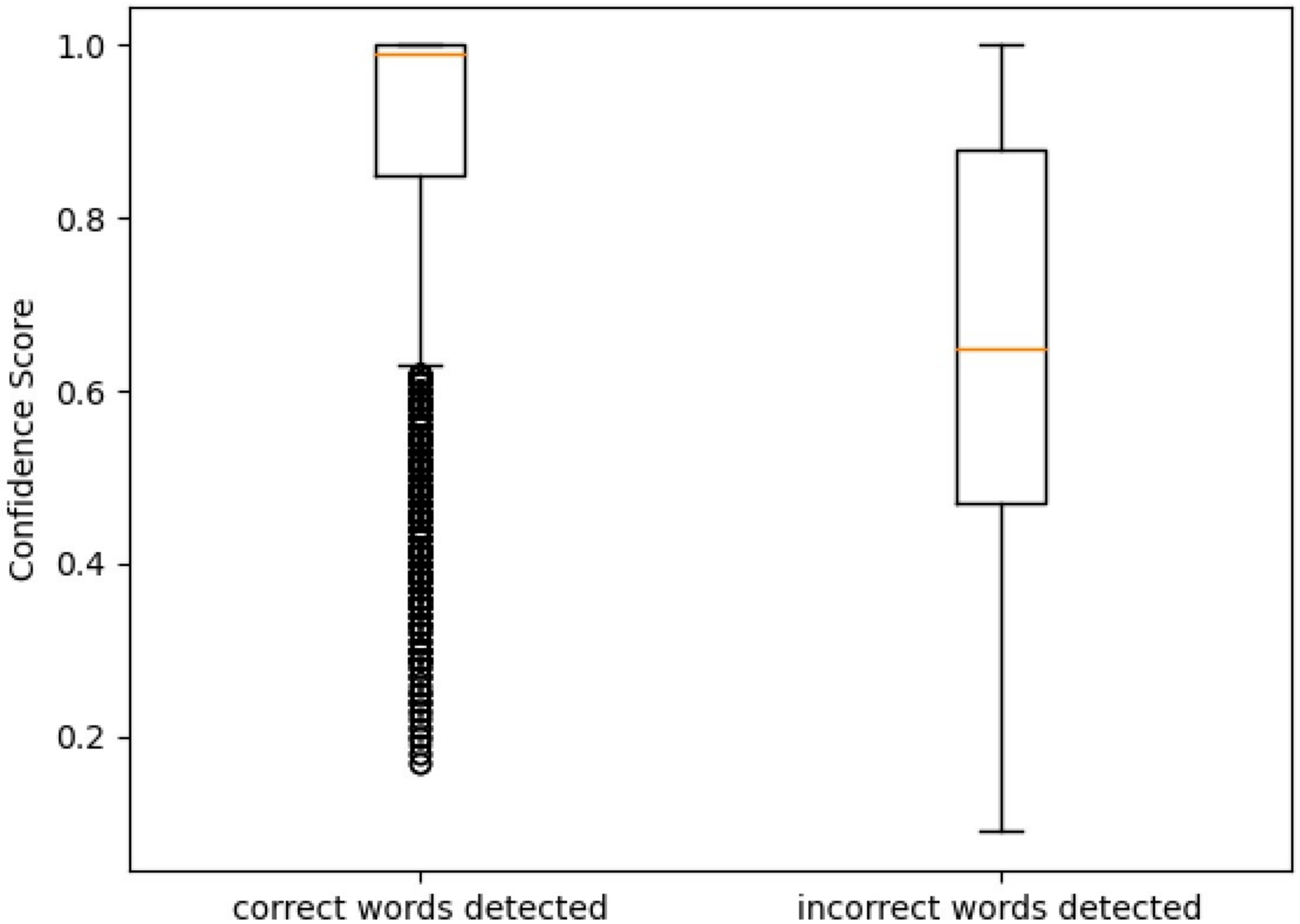}
\caption{Word confidence score analysis over the revised segmented version of the three test sets.}

\label{word conf}
\end{figure}
The confidence in prediction of a word can be a useful indicator for the purpose of assessing the quality of the predicted output, and can be used for practical applications such as song retrieval through lyrics. Kaldi provides a confidence score in prediction using minimum Bayes risk decoding \footnote{\url{https://github.com/kaldi-asr/kaldi/blob/master/egs/wsj/s5/steps/conf/get_ctm_conf.sh}}. The word confidence analysis is conducted on the results of the segmented test sets with the proposed music-robust approach. As shown in Fig.~\ref{word conf}, in general for the correctly recognized words, the confidence score is high, and the words that are incorrectly recognized show lower confidence score. This shows that when the system predicts with high confidence, the predicted words are more likely to be correct than incorrect.

\subsection{Comparison with Prior Studies}

We compare the proposed music-robust model with the existing approaches in Table~\ref{existing}. Specifically, we compare our proposed model with the systems~\cite{dabikesheffield,demirel2020recursive,gaolyrics} submitted to the lyrics transcription competition in the 16th Music Information Retrieval Evaluation eXchange International Benchmarking Competition (MIREX 2020), which is a well-known lyrics transcription challenge and the results are publicly available \footnote{\url{https://www.music-ir.org/mirex/wiki/2020:Lyrics_Transcription_Results}}. Since the song "clock" in Hansen datasets was cut half from the original audio for MIREX 2020 competition, we exclude the song "clock" from Hansen dataset and mark it as Hansen-ex in Table~\ref{existing} for fairer comparison. ~\cite{gaolyrics} consider only music-present features while \cite{dabikesheffield,demirel2020recursive} only use music-removed features for AM training, whereas our proposed music-robust system benefits from both music-present and music-removed features.

As can be found in Table~\ref{existing}, our proposed music-robust system shows considerable improvement for the three published test datasets in lyrics transcription performance compared to all the previous studies. We note that our model also shows 8-9\% improvement over the recent top results GGL1~\cite{gaolyrics} in MIREX 2020.

We expect that transcription performance over shorter segments would be better than over the whole song, because in Viterbi decoding for long utterances, the errors tend to accumulate.
Moreover, speech recognition~\cite{tu2018hybrid,tu2019dnn,hermus2006review} and lyrics transcription of monophonic music~\cite{dabikesheffield,demirel2020recursive} are generally performed on segmented data.
Therefore, we segment the songs in the three test sets manually into short utterances of 20-30 seconds, which is released for the development of the research community\footnote{\url{https://github.com/xiaoxue1117/ALTP_chords_lyrics}}. We also further clean-up the reference lyrics with minor transcription corrections. We present the lyrics recognition performance on the revised segmented test set and compare it with the original whole song data in Table~\ref{segment}. 
We found that the revised segmented data performs consistently better than the whole song version across the baseline models and the proposed model, as expected. We will publish our segmented test sets for the research community upon paper acceptance.

\section{Conclusions}
\label{Conclusion}

This paper presents a novel music-robust automatic lyrics transcription approach via building music-robust features for lyrics acoustic modeling and interpolated language modeling of polyphonic music. The combination of music-present features with music-removed features offers compensation for the missed vocal components caused by singing source separation techniques, thereby conveying complementary information about the vocals while also emphasizing on the singing vocals that are highly related to lyrics intelligibility. We also investigate the interpolated LM to bridge the in-domain linguistic peculiarities with high resource out-domain information to enhance lyrics transcription performance. The experimental results show the proposed music-robust model is superior over the existing approaches on publicly available test sets.

\section{Acknowledgement}
This research is supported by the Agency for Science, Technology and Research (A*STAR) under its AME Programmatic Funding Scheme (Project No. A18A2b0046). This work is supported by A*STAR under its RIE2020 Advanced Manufacturing and Engineering Domain (AME) Programmatic Grant (Grant No. A1687b0033, Project Title: Spiking Neural Networks). This research work is also supported by Academic Research Council, Ministry of Education (ARC, MOE), Singapore. Grant: MOE2018-T2-2-127. Title: Learning Generative and Parameterized Interactive Sequence Models with RNNs.

\bibliography{smc2022bib}

\end{document}